\documentclass[aps,prl,preprint,groupedaddress]{revtex4}

\def\av#1{\left\langle#1\right\rangle}

\usepackage{graphicx}
\usepackage{bm}
\usepackage{graphicx}
\usepackage{subfigure}
\usepackage{latexsym}
\usepackage{amstext,amssymb,amsfonts}
\usepackage{epsfig}
\usepackage{color}

\begin{document}

\title{Anomalous biased diffusion in networks}

\author{Loukas Skarpalezos}

\affiliation{Department of Physics, University of Thessaloniki, 54124 Thessaloniki, Greece}

\author{Aristotelis Kittas}

\affiliation{Department of Physics, University of Thessaloniki, 54124 Thessaloniki, Greece}

\author{Panos Argyrakis}

\affiliation{Department of Physics, University of Thessaloniki, 54124 Thessaloniki, Greece}

\author{Reuven Cohen}

\affiliation{Department of Mathematics, Bar-Ilan University, Ramat-Gan 52900, Israel}

\author{Shlomo Havlin}

\affiliation{Department of Physics, Bar-Ilan University, Ramat-Gan 52900, Israel}

\date{\today}

\begin{abstract}
We study diffusion with a bias towards a target node in networks. This problem is relevant to efficient routing strategies in emerging communication networks like optical networks. Bias is represented by a probability $p$ of the packet/particle to travel at every hop towards a site which is along  the shortest path to the target node. We investigate the scaling of the mean first passage time (MFPT) with the size of the network. We find by using theoretical analysis and computer simulations that for Random Regular (RR) and Erd\H{o}s-R\'{e}nyi (ER) networks, there exists a threshold probability, $p_{th}$, such that for $p<p_{th}$ the MFPT scales anomalously as  $N^\alpha$, where $N$ is the number of nodes, and $\alpha$ depends on $p$. For $p>p_{th}$ the MFPT scales logarithmically with $N$. The threshold value $p_{th}$ of the bias parameter for which the regime transition occurs is found to depend only on the mean degree of the nodes. An exact solution for every value of $p$ is given for the scaling of the MFPT in RR networks. The regime transition is also observed for the second moment of the probability distribution function, the standard deviation. 
\end{abstract}

\maketitle

Recently there has been growing interest in investigating the properties of complex networks \cite{barabasi_sci286,albert_rev74,newman_siam45,dorogovtsev_adv51,caldarelli2012networks}. These include systems from markedly different disciplines, as communication networks, the Internet itself, social networks, networks of collaboration between scientists, transport networks, gene regulatory networks, and many other examples in biology, sociology, economics and even linguistics, with new systems being added continuously to the list \cite{Pastor-Satorras:2004:ESI:1076357,DorogovtsevBook,Cohen_Havlin:2010,newman_book,Barrat:2008:DPC:1521587,Bashan_natcom}.

Sending messages through a network in the form of packets in an efficient way is one of the most challenging problems in today's communication technologies. It is obvious that a fully biased walk (with probability to stay on the shortest path equal to 1) would be the most efficient way to send a message, if the exact structure of the network is known. But, quite often, as in the case of wireless sensor networks  \cite{avin_query}, ad-hoc networks \cite{yossef_mobi} and peer-to-peer networks \cite{gkantsidis_peer}, due to the continuously dynamically changing infrastructure, the application of routing tables is not possible, and the so called hot potato/random walk routing protocol is preferable, because it can naturally cope with failures or disconnections of nodes. The problem with such a procedure, in which data packets traverse the network in a random fashion, is a significant increase of the hitting time. For this reason, new protocols have been proposed recently \cite{beraldi1}, that are based on the idea of biased random walks  and which can significantly reduce the hitting time in such networks. For example, the Lukewarm Potato Protocol \cite{beraldi2}, is totally tunable (with the value of just one threshold parameter) and can operate anywhere in the continuum from the hot potato/random walk forwarding protocol to a deterministic shortest path forwarding protocol. But also, in a more general manner, we can consider that every routing protocol, which uses deflection (hot potato) routing in certain circumstances (e.g. insufficient storage space of the node or a disconnected node), can be represented by a biased random walk process since it uses the shortest path only if it is possible. This problem is also very relevant in optical networks where optical switches pay a large price for packet storing (with the conversion of light to elecronic signals). The result is a limited storing capacity of optical switches that must route packages in a random direction in the case the destination path is overloaded or they have reached the storage limit. Therefore, the probabilty to stay on the shortest path may, in certain cases, have a small value (optical switches with unsufficient storage capacity), and in other cases, a large one (few disfunctioning nodes). It is consequently of great interest, and it is the subject of this work, to understand how the diffusion process is affected when a tunable bias along the shortest path is used and to theoretically study the scaling properties of such biased diffusion processes.

Random Regular (RR) networks are networks where all nodes have exactly the same number of edges (connections). They constitute a well studied mathematical model which is suitable for exact analysis of its properties. The Erd\H{o}s-R\'{e}nyi (ER) model \cite{erdos_publ6,erdos_1960,Bollobas_RandGraph} is a well known simple model, which generates random graphs by setting an edge between each pair of nodes with a probability $q$, independently of the other edges. This yields (in the limit $N\rightarrow\infty$) a Poisson distribution (for $q<1$) of the node degree $k$: $P(k)=\frac{\av{k}^k}{k!}e^{-\av{k}}$ with $\av{k}=q(N-1)$, with  $q=1$ giving the completely connected graph. 
An important property of networks is the average path length $D$ between two nodes. For the case of RR and ER networks it has been shown that $D$ scales as $\ln N $. This dependence is the origin of the well known small world phenomena in networks.
 
Random walks have interesting properties which may depend on the dimension and the structure of the medium in which they are confined \cite{avraham_diffusion,weiss_random,redner_mfpt,havlin_adv36,gallos_prl}, e.g. lattices or complex networks. Diffusion is a very natural mode of transport, where hopping from one node to the next is unaffected by the history of the walk \cite{weiss_random,bollt_njphys7}. A measure of diffusion which has been extensively studied (see e.g. \cite{redner_mfpt,condamin_nat450,condamin_prl95,baronchelli_pre73,sood_prl99,sood_jphys38,argy_physa363}), is the first-passage time (FPT), which is the time required for a random walker to reach a given target point for the first time. The importance of FPT originates from the crucial role played by first encounter properties in various real situations, including transport in disordered media, neuron firing dynamics, buying/selling on the stock market, spreading of diseases or target search processes \cite{redner_mfpt, condamin_nat450}. 

The properties of the first-passage time have been investigated in a variety of networks. Baronchelli and Loreto \cite{baronchelli_pre73}, using the concept of rings, have shown that the FPT probability distribution in ER networks is an exponential decay and FPT vs the degree of the target node is a power law for various networks, such as ER networks, the Barab\'{a}si-Albert model (BA) \cite{barabasi99emergence}, as well as the Internet. An analytical formula has also been derived for the mean first-passage time (MFPT) of a random walker from one node  to another, namely $\av{T_{ij}}$ (mean transit time), on networks  \cite{noh_prl92}. Note that in this case a random walk motion from node $i$ to $j$ is not symmetric with the motion in the opposite direction. The size scaling of $\av{T_{ij}}$ has been studied in a variety of systems and geometries \cite{bollt_njphys7}. The trapping problem on networks which is closely related to MFPT was studied by Kittas et al \cite{kittas08}. Biased random walks on networks have also been studied \cite{sood_prl99}, including local navigation rules (see e.g. \cite{fronczak_pre80,adamic_pre64,tadic_physa332,wang_pre73}).

\begin{figure}[htbp]
    \begin{center}
\includegraphics[width=6.in]{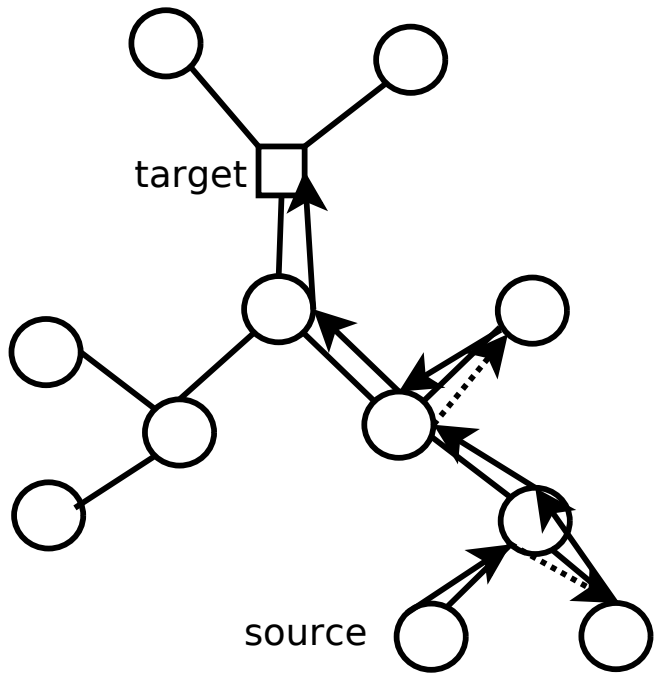}
    \end{center}
     \caption{ Illustration of the biased diffusion process. The arrows represent moves. Full arrows represents movement along the shortest path, while dashed arrows represent random steps. The destination node is represented by a square.}
\label{fig1}
\end{figure}

We use Monte Carlo computer simulations implemented by the following algorithm: Initially, a source and a target node are selected at random. The particle travels from the source to the target node either randomly, or with a bias (for a schematic see Fig. \ref{fig1}). The bias is expressed by a parameter $p$, which is the probability that the particle at each time step travels towards the target node using the shortest path to it. To calculate the shortest path we use the Breadth-First-Search (BFS) algorithm as described in \cite{cormen_algorithms}. Given a graph $G=(V, E)$ and a specific source vertex $s$, BFS systematically explores the edges of $G$ to record every vertex that is reachable from $s$. It computes the distance from $s$ to each reachable vertex, which is the smallest number of edges. We use the target node as the BFS "source" $s$ and identify the geodesic distances from the target to every node in the network, i.e. the number of links in the shortest path from the target to any arbitrary node. Thus, each node is assigned a number, which indicates its distance from the target. When the particle moves, it jumps to one of its adjacent nodes, which belong to the shortest path with probability $p$, or to a random node (including the ones in the shortest path) with probability $1-p$. Consequently, for $p=1$ the particle always travels on the shortest path, while for $p=0$ it performs a stochastic random walk. We consider the process only on the largest cluster of the network (also discovered with the BFS algorithm). We perform $10^5$ total runs (1000 networks, considering 100 pairs of random source-target nodes for each network realization).

\begin{figure}[htbp]
    \begin{center}
    \subfigure[]{
    \includegraphics[width=3.in]{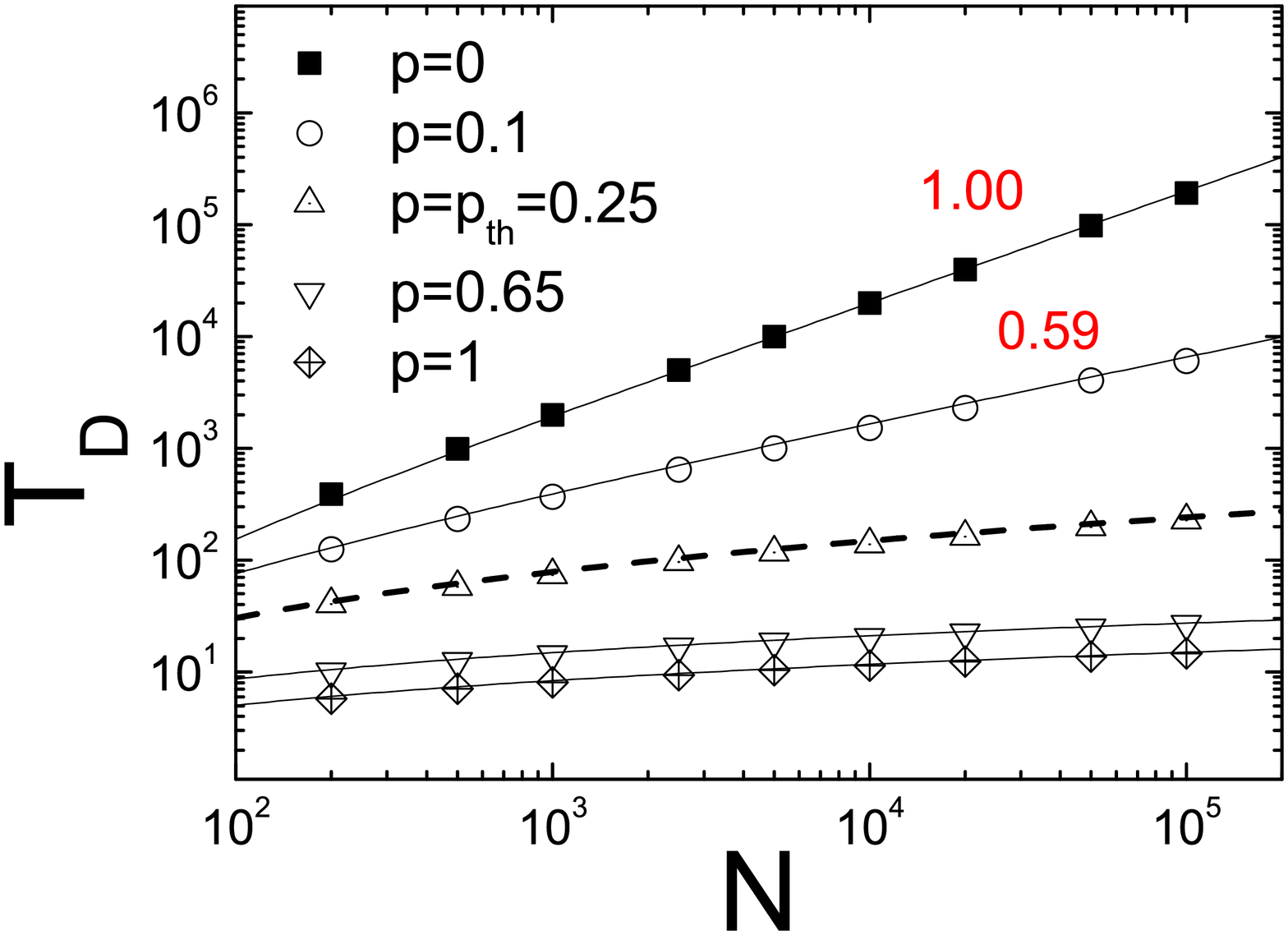}
    \label{fig2:a}}
    \subfigure[]{
    \includegraphics[width=3.in]{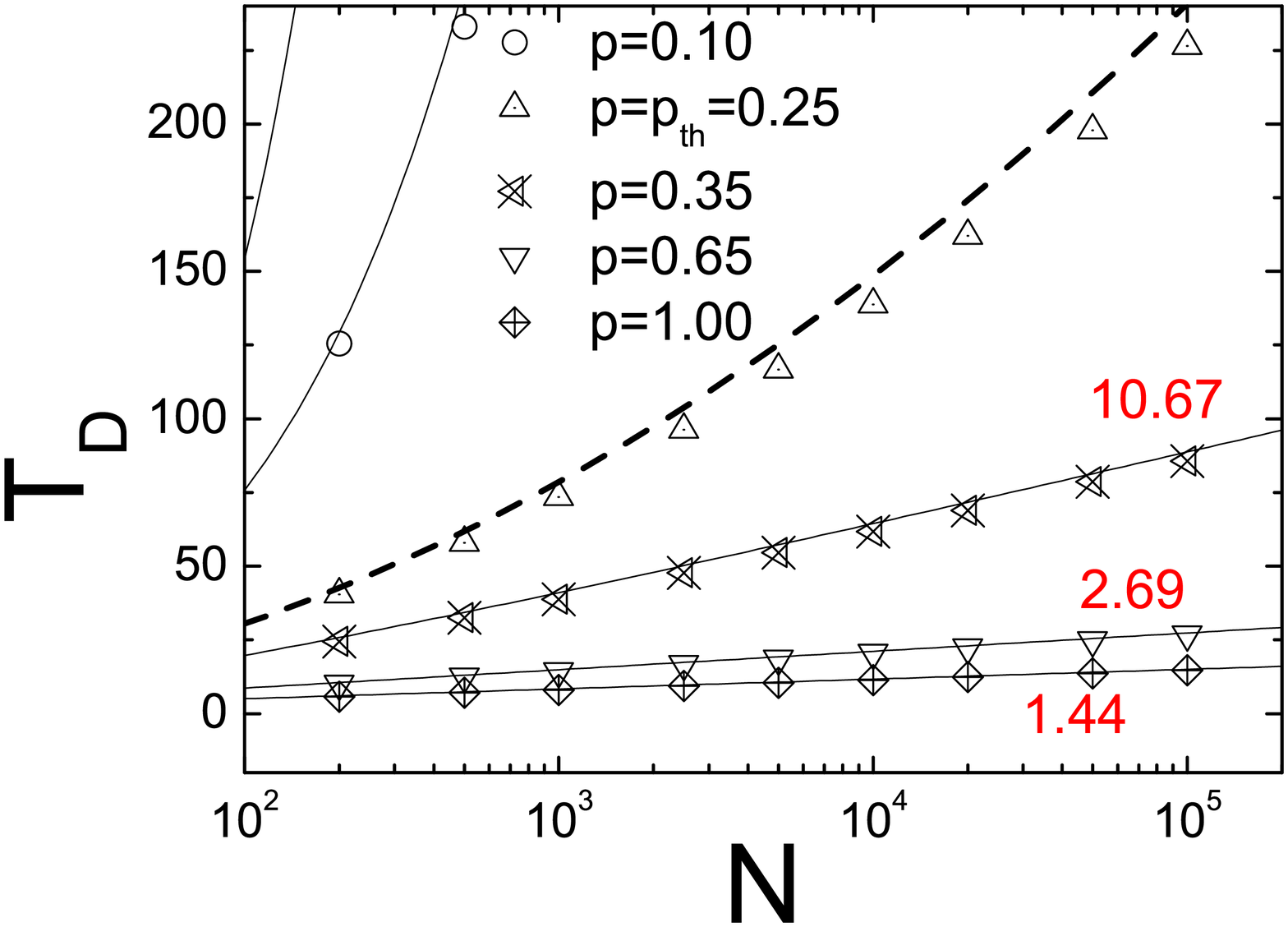}
    \label{fig2:b}}
    \subfigure[]{
    \includegraphics[width=3.in]{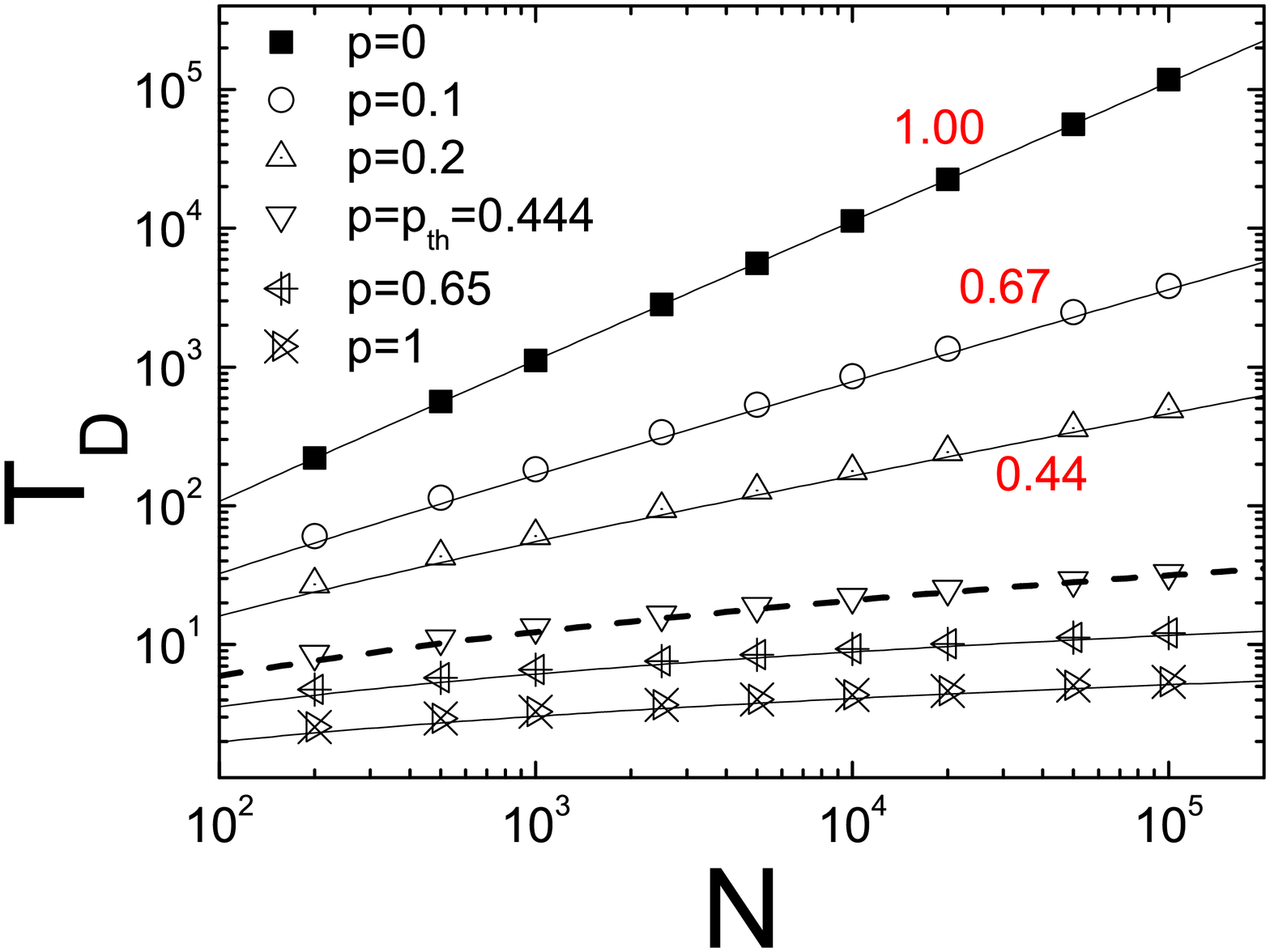}
    \label{fig2:c}}
    \subfigure[]{
    \includegraphics[width=3.in]{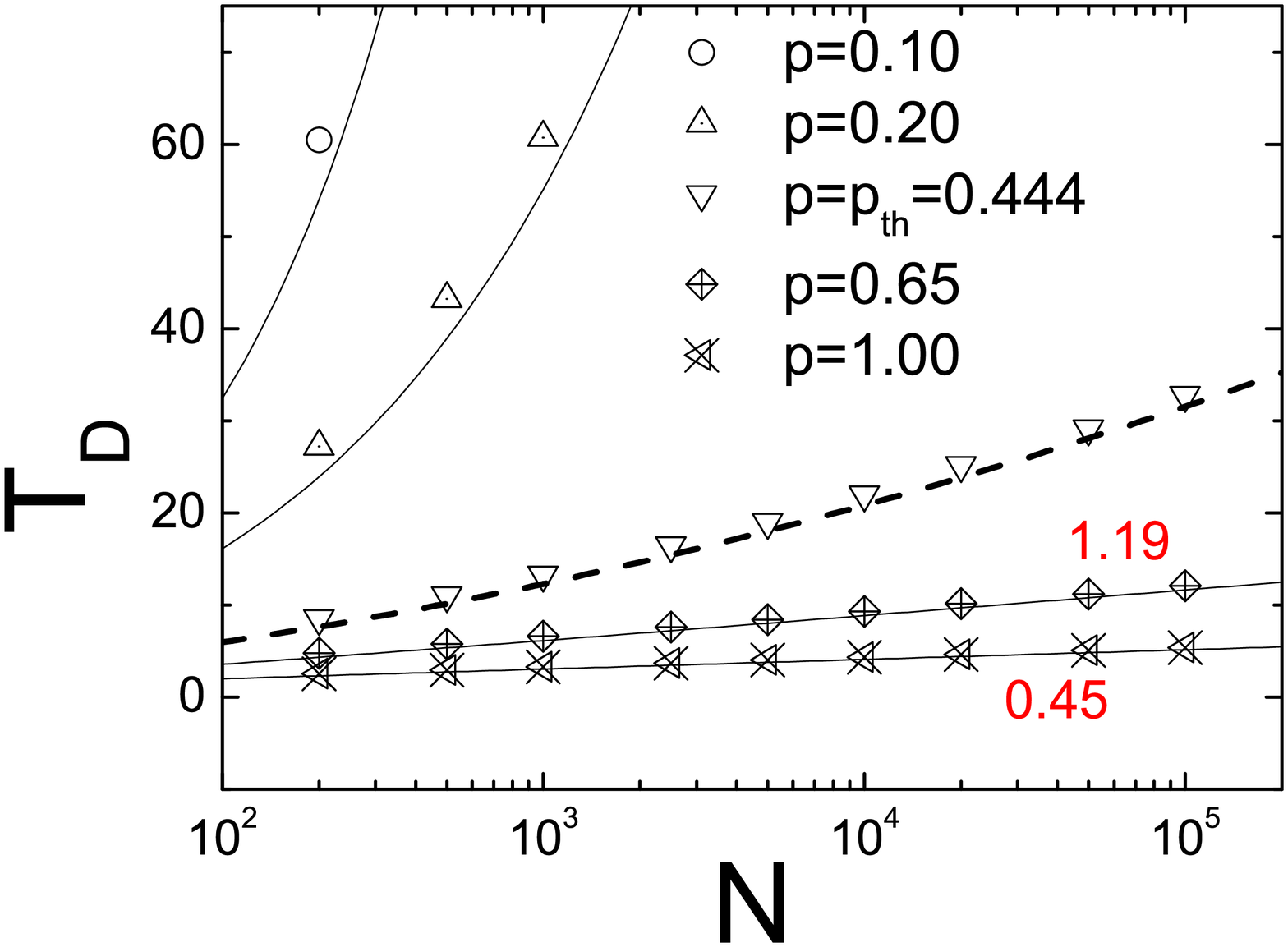}
    \label{fig2:d}}
    \end{center}
     \caption{(Color online) (a) Log-log and (b) Log-linear representation of MFPT ($T_D$) vs $N$ for RR networks with $k=3$, and (c) Log-log and (d) Log-linear representation of MFPT vs $N$ for RR networks with $k=10$ for various values of $p$. Solid lines represent analytical solutions and dashed lines correspond to the threshold value, where $T_D\sim D^2$, i.e., regular diffusion. The measured slopes (red) of the power law regime and the prefactors (red) of the logarithmic regime are in excellent agreement with the values given by (\ref{eq:slope}) and (\ref{eq:log}), respectively.}
\label{fig2}
\end{figure}

Firstly, we investigate the scaling of the MFPT with system size $N$ (number of nodes of the network) for RR networks. We find that the value of $p$ has a large effect on the scaling of MFPT, with one range of large $p$ having  a logarithmic scaling and another of small $p$ having a power law function of $N$ (see Fig. \ref{fig2}). As $p$ increases the system size becomes less relevant and the MFPT scales logarithmically with the system size, similar to the diameter of the network.

For the analytical approach of the case of RR networks, we consider a walk on a finite tree of depth $D$ with reflecting boundary conditions at the leaves (ends). We go towards the root with probability $p$ and hop to a random neighbor with probability $1-p$. Since there are $k$ neighbors to each node, there is a probability $(1-p)/k$ that we may choose the link going towards the root. Eventually, this can be mapped to a random walk on a finite segment $\{0,1,\ldots,D\}$. Since the number of nodes at a distance $d$ from the source is approximately $n_d=k(k-1)^{d-1}$, and the total number of nodes is
\begin{equation}
 N=1+\sum_{i=1}^D k(k-1)^{d-1}=1+k\frac{(k-1)^D-1}{k-2}\;,
 \end{equation}
 it follows that
\begin{equation} 
D=\frac{\log\left(1+(k-2)(N-1)/k\right)}{\log (k-1)}\approx\frac{\log\left((k-2)N/k\right)}{\log (k-1)}
\end{equation}
 is the average distance and the probability of going towards the target 
is $p'=p+(1-p)/k$.
Denote by $T_i$ the average time it takes the walker to reach the destination 
when it is at distance $i$ from it.
The recurrence equations are 
\begin{equation}
\label{eq:recur}
T_i=1+p'T_{i-1}+(1-p')T_{i+1}\;,
\end{equation}
for $0<i<D$ and
\begin{eqnarray}
\label{eq:recur_t0}
T_0&=&0\;,\\
\label{eq:recur_tD}
T_D&=&1+p'T_{D-1}+(1-p')T_D\;.
\end{eqnarray}
The solution of Eq. (\ref{eq:recur}) is 
\begin{equation}
T_i=\frac{i}{2p'-1}+c_1+c_2\left(\frac{p'}{1-p'}\right)^i\;.
\end{equation}
Substituting in Eq. (\ref{eq:recur_t0}) and (\ref{eq:recur_tD}) one obtains
\begin{equation}
c_1=-c_2=\frac{1-p'}{(2p'-1)^2}\left(\frac{1-p'}{p'}\right)^D
\end{equation}
Thus,
\begin{equation}
\label{eq:sol}
T_D=\frac{D}{2p'-1}+\frac{1-p'}{(2p'-1)^2}\left[\left(\frac{1-p'}{p'}\right)^D-1\right]\;.
\end{equation}
A better approximation is obtained when taking into consideration the probability of selecting a pair of nodes at distance $i$ from each other. The probability of choosing such a pair is approximately 
\begin{equation}
P(i)=\frac{k(k-1)^{i-1}}{\sum_{i=1}^D k(k-1)^{i-1}}\;.
\end{equation}
Thus, the expected time is
\begin{eqnarray}
E[T]=\sum_{i=1}^D P(i)T_i=\frac{D(k-1)^{D+1}-(D+1)(k-1)^D+1}{(k-2)((k-1)^D-1)}+c_1-\nonumber\\
c_1\frac{(k-1)\frac{p'}{1-p'}\left(\left((k-1)\frac{p'}{1-p'}\right)^D-1\right)}{\left((k-1)\frac{p'}{1-p'}-1\right)
\left((k-1)^D-1)\right)}\;.
\end{eqnarray}

Therefore, if $p'>1/2$ (i.e. $p>(k-2)/(2k-2)$), the first term of Eq.(\ref{eq:sol}) dominates and we have
that the first passage time is approximately 
\begin{equation}
\label{eq:log}
T_D\approx\frac{D}{2p'-1}\approx \frac{\log\left((k-2)N/k\right)}{(2p+2(1-p)/k-1)\log(k-1)}\;.
\end{equation}
Whereas, if $p'<1/2$ the second term dominates and we have 
\begin{equation}
\label{eq:pow}
T_D\approx\frac{1-p'}{(2p'-1)^2}\left(\frac{1-p'}{p'}\right)^D
\approx \frac{1-p'}{(2p'-1)^2}\left(\frac{1-p'}{p'}\right)^{\log\left((k-2)N/k\right)/\log(k-1)}\propto N^\alpha\;,
\end{equation}
where
\begin{equation}
\label{eq:slope}
\alpha= \frac{\log\frac{1-p'}{p'}}{\log(k-1)}\;.
\end{equation}
The minimum value for $p'$ is $p'=1/k$ (obtained for $p=0$). In this case $1-p'=(k-1)/k$ and $\alpha=1$, i.e., on average the walk moves randomly with no preferred direction and reaches a large fraction of the nodes in the network before reaching the target, as expected.

For the case $p'=1/2$ the solution for the equations
becomes 
\begin{equation}
T_i=(2D+1)i-i^2\;,
\end{equation}
and therefore,
\begin{equation}
T_D=D(D+1)=\frac{\log\left((k-2)N/k\right)}{\log(k-1)}\left(\frac{\log\left((k-2)N/k\right)}{\log(k-1)}+1\right)\;.
\end{equation}
Thus, it behaves like normal diffusion, where the time needed to reach distance 
$D$ is of the order $D^2$ \cite{weiss_random,redner_mfpt,avraham_diffusion}.

From the above analysis we clearly see that for RR networks there exists an abrupt change from a power law behavior to logarithmic dependence on  $N$ for the  MFPT. The limit between these two radically different scaling behaviors corresponds to the threshold value of the bias parameter $p_{th}=(k-2)/(2k-2)$. In Fig. \ref{fig2} we compare the analytical solution (\ref{eq:sol}) with the results of the Monte Carlo simulations for $k=3$ and $k=10$. The measured slopes of the power law regime and the prefactors of the logarithmic regime are in excellent agreement with the values given by (\ref{eq:slope}) and (\ref{eq:log}), respectively. 

\begin{figure}[htbp]
    \begin{center}
    \subfigure[]{
    \includegraphics[width=3.in]{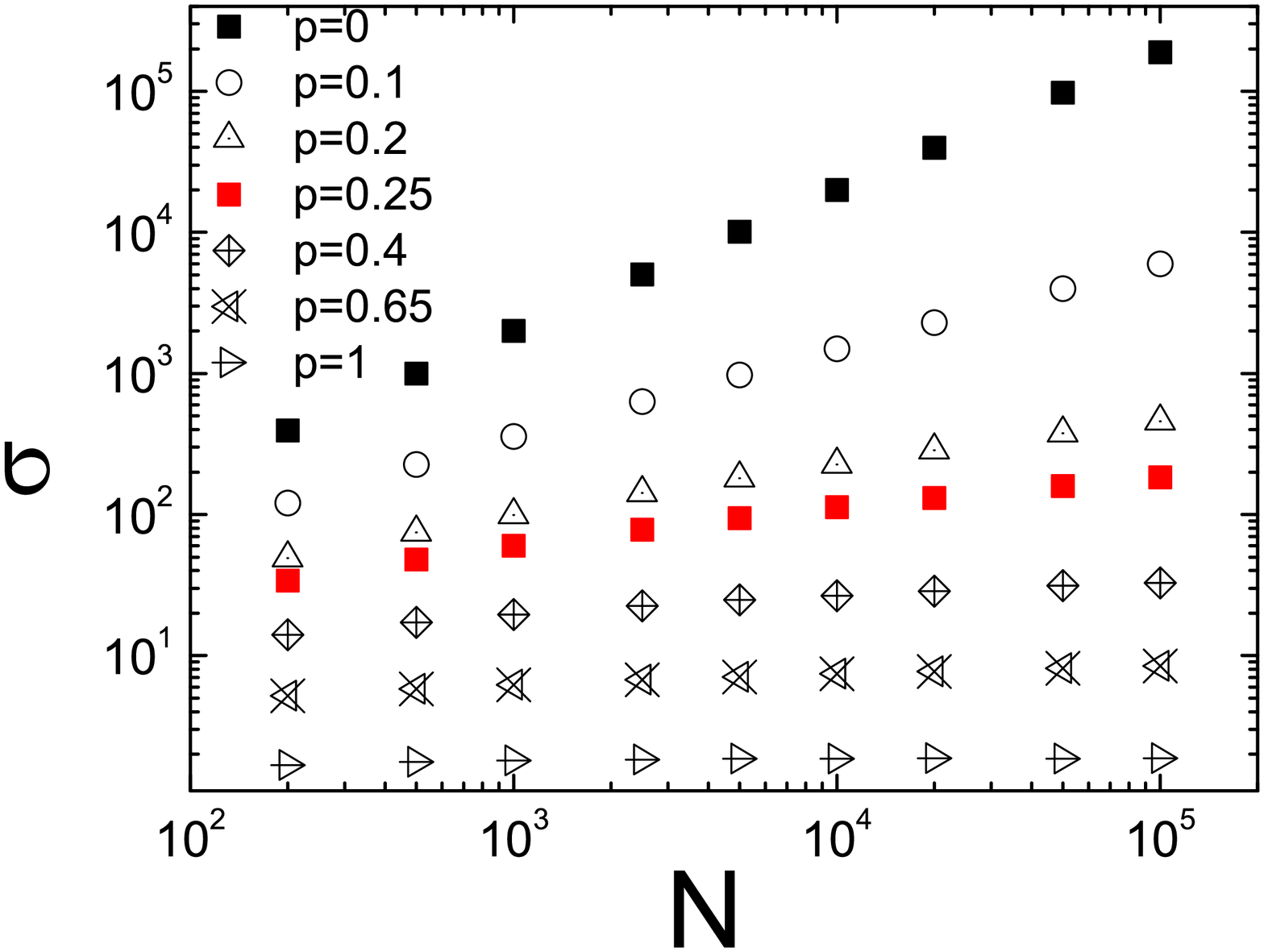}
    \label{fig3:a}}
    \subfigure[]{
    \includegraphics[width=3.in]{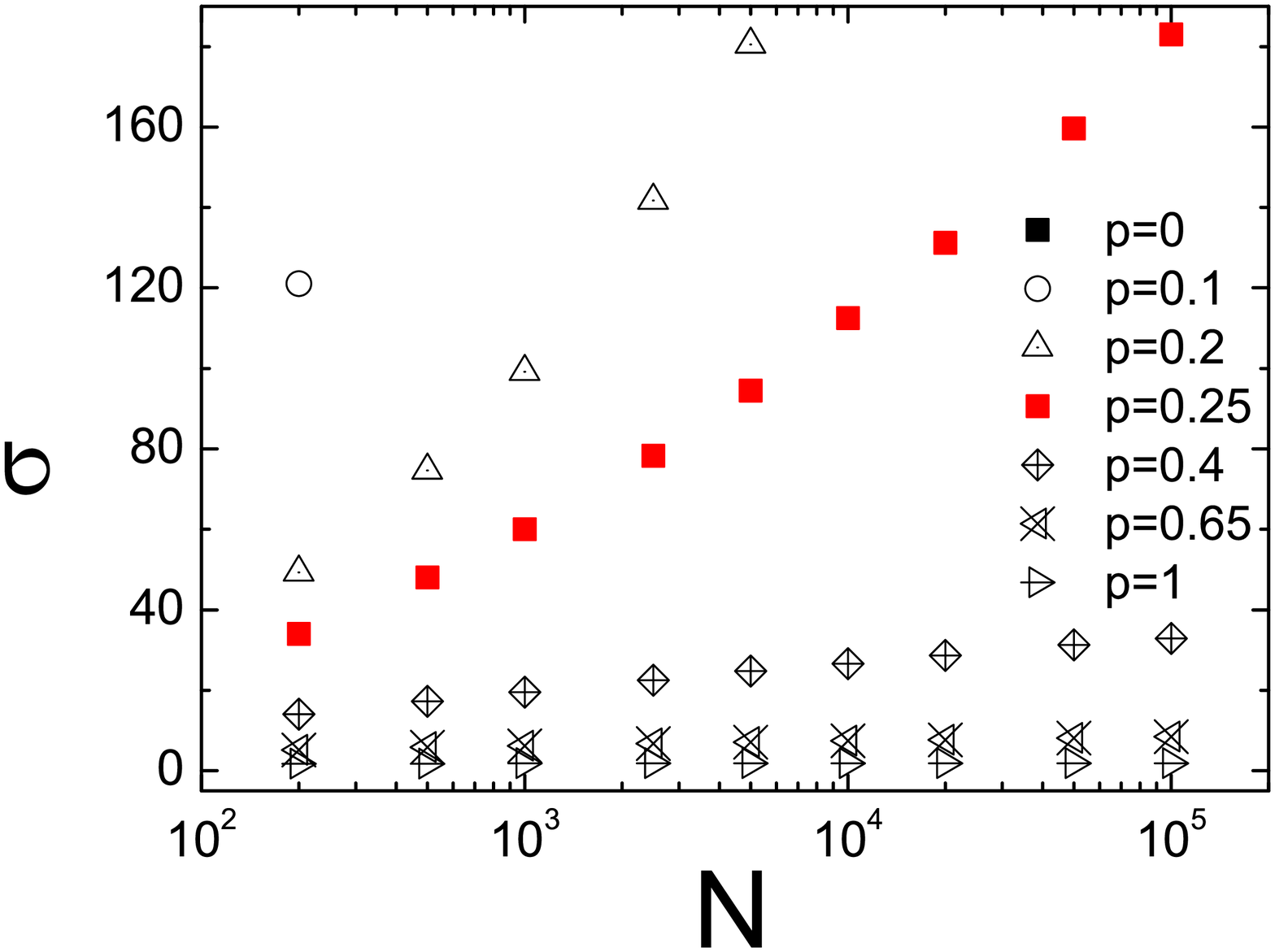}
    \label{fig3:b}}
    \subfigure[]{
    \includegraphics[width=3.in]{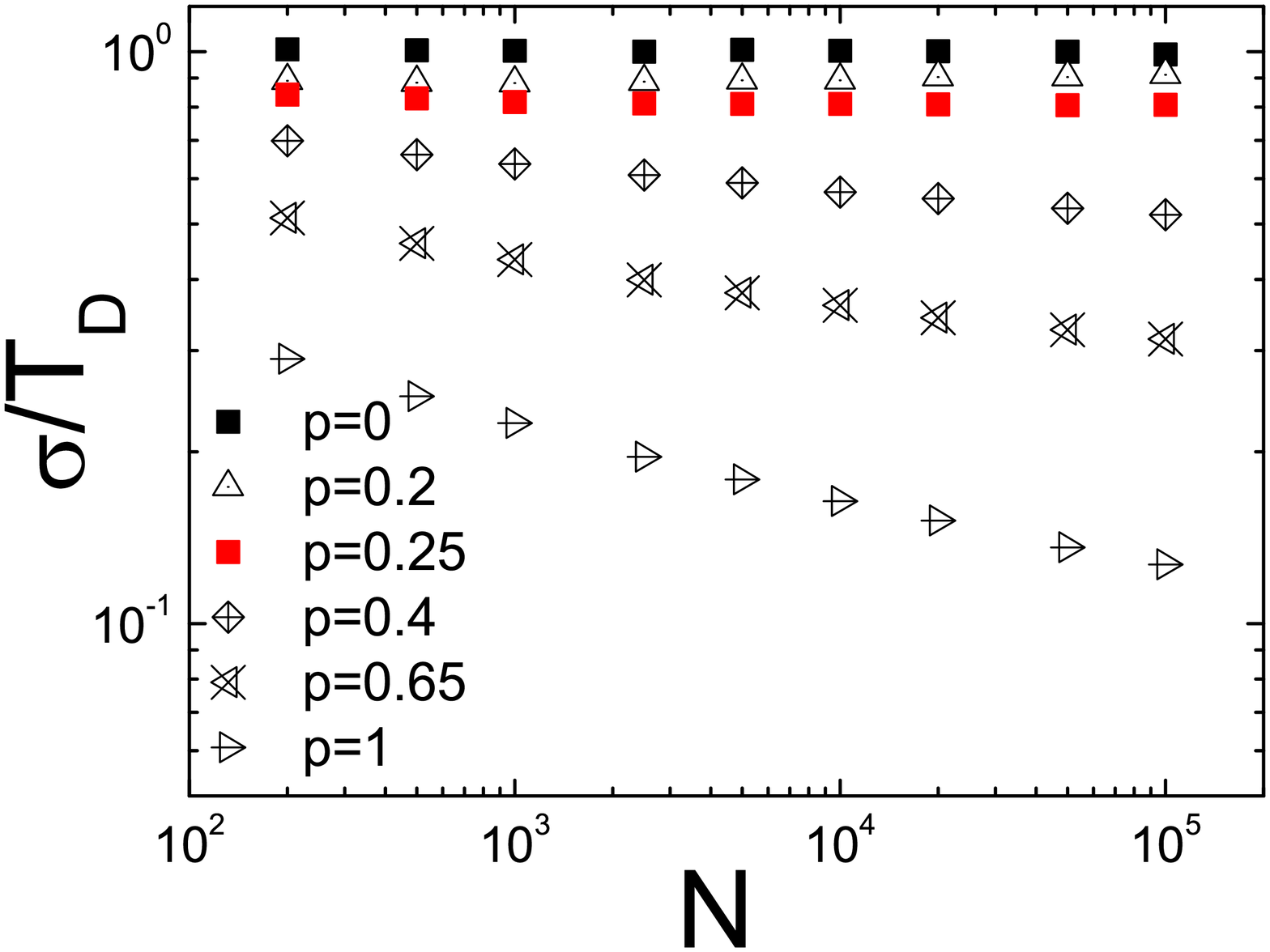}
    \label{fig3:c}}
    \subfigure[]{
    \includegraphics[width=3.in]{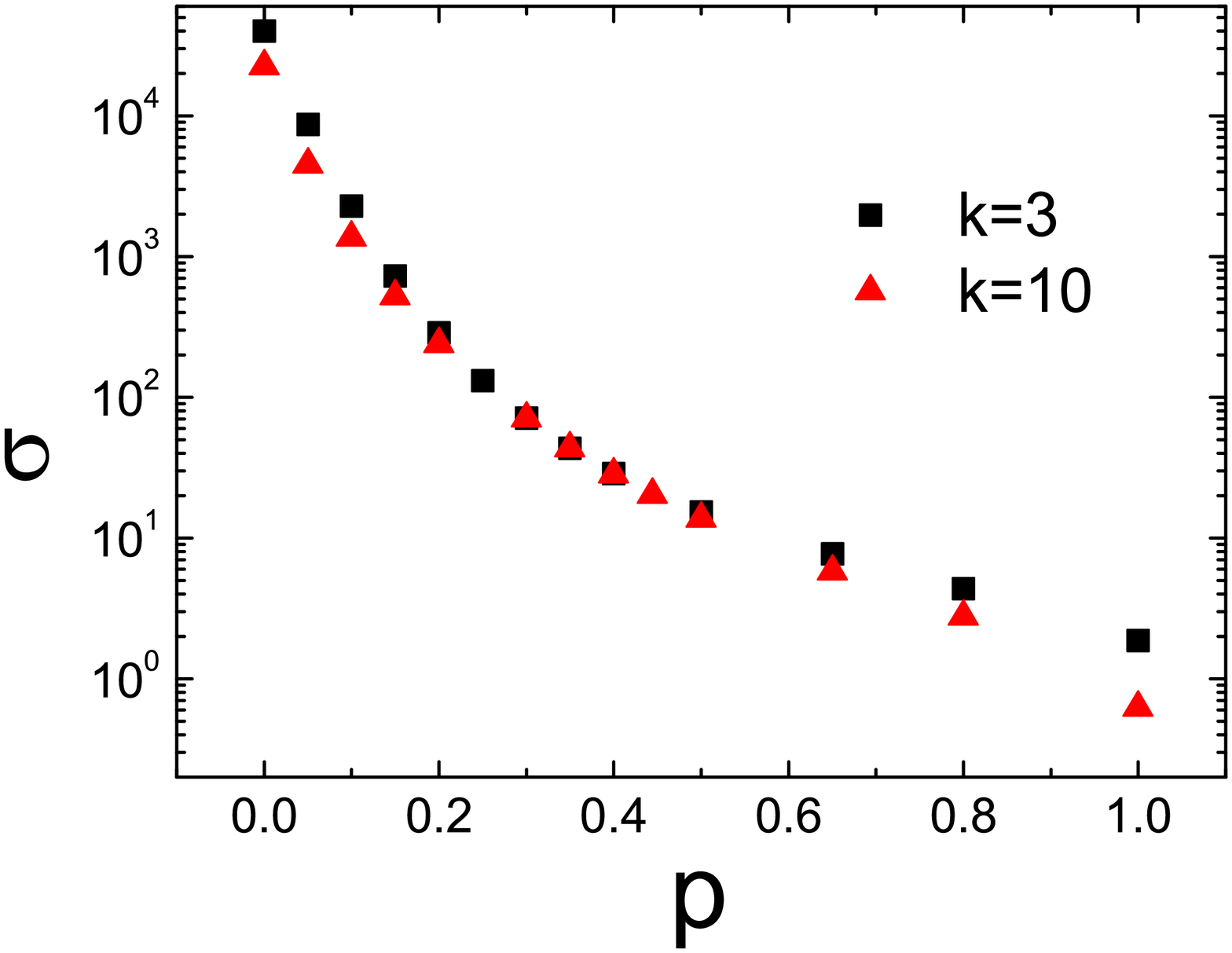}
    \label{fig3:d}}
    \end{center}
     \caption{(a) Log-log and (b) Log-linear plot of the standard deviation $\sigma$ vs $N$ for RR networks with $k=3$, (c) plot of $\sigma /T_D$ vs $N$ for RR networks with $k=3$ and (d) plot of $\sigma$ versus the value of the bias parameter $p$ for $N$=20000.}
\label{fig3}
\end{figure}

In Fig. \ref{fig3} we investigate the  behavior of the standard deviation $\sigma$, which is the second moment of the probability distribution function of the FPT for RR networks. In Fig. \ref{fig3:a} and \ref{fig3:b} we see that the scaling of the standard deviation with the size of the network largely resembles that of the MFPT, with two different regimes separated by the threshold value $p_{th}$. This resemblance and the existance of the regime transition for the same threshold value $p_{th}$ is made more clear in  \ref{fig3:c} where the scaling of  the ratio $\sigma$/MFPT is represented. We see that for $p<p_{th}$, the scaling of the two quantities is the same, while for $p>p_{th}$, the standard deviation scales slower with $N$ than the MFPT. In Fig. \ref{fig3:d} we see the dependence of the standard deviation on the value of $p$ for a fixed network size. We see that the standard deviation decreases to reach a very small value for a fully biased diffusion. This is expected since for a fully biased walk the probability distribution function corresponds to a delta function.

\begin{figure}[htbp]
    \begin{center}
    \subfigure[]{
    \includegraphics[width=3.in]{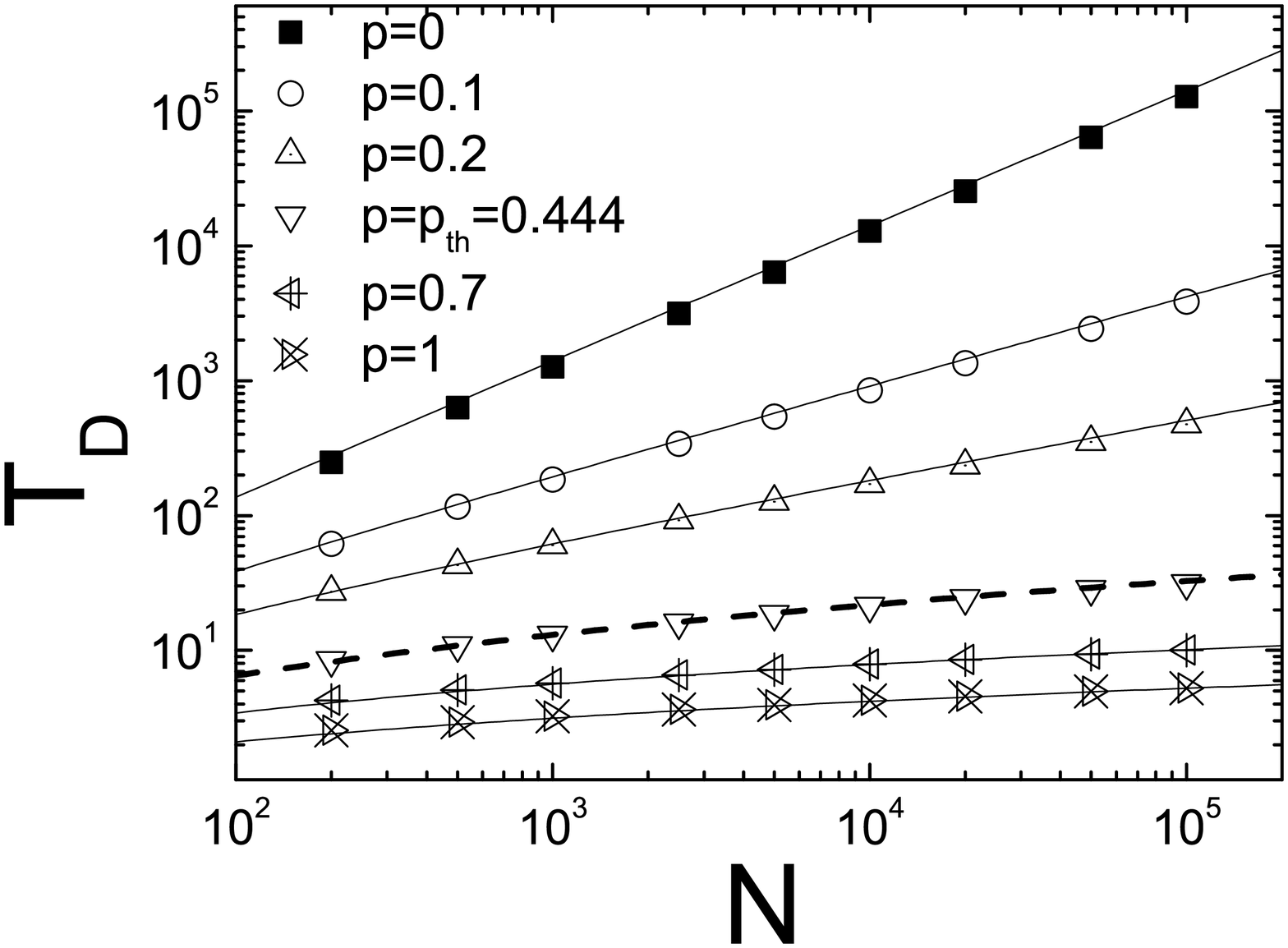}
    \label{fig4:a}}
    \subfigure[]{
    \includegraphics[width=3.in]{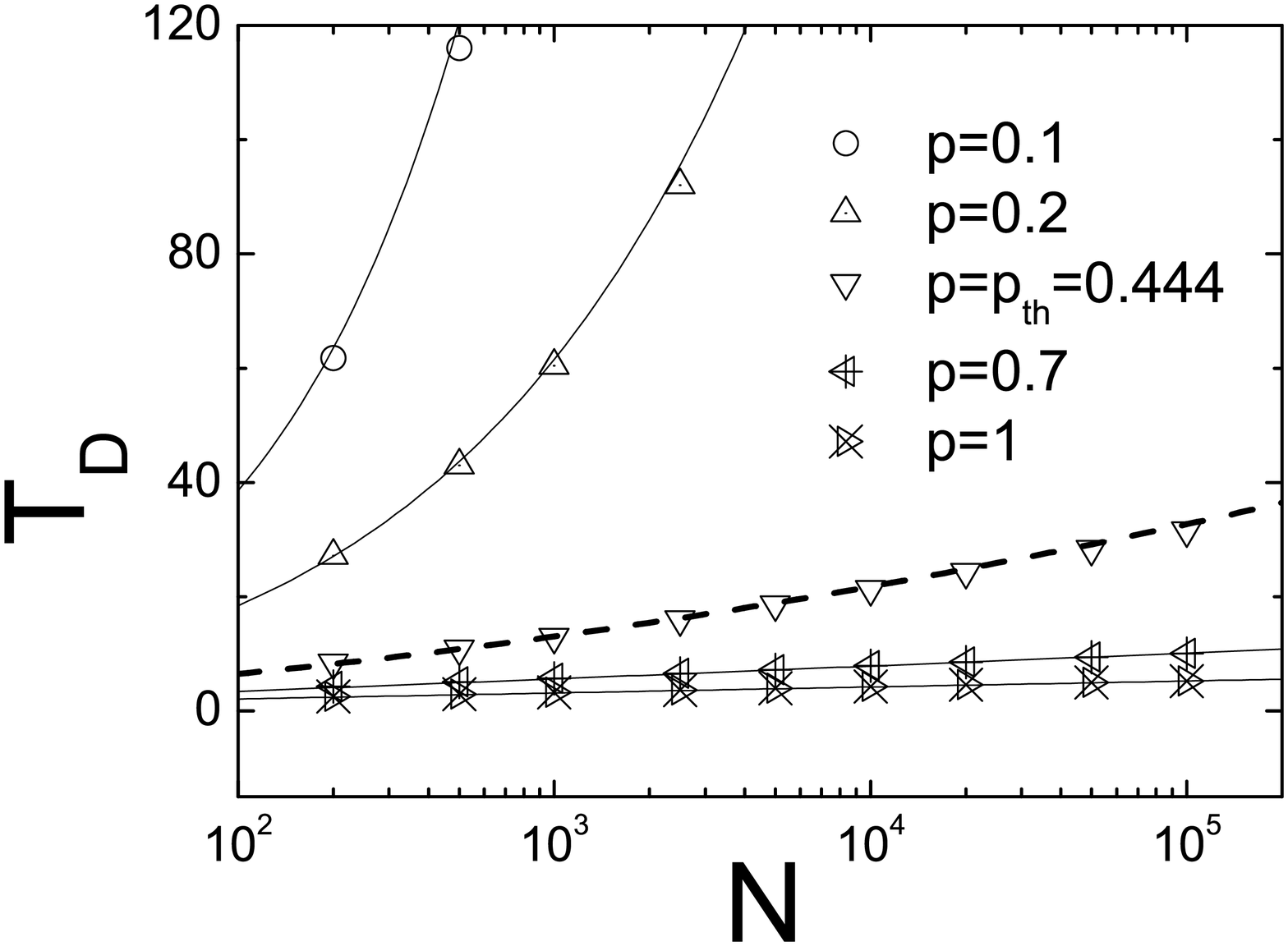}
    \label{fig4:b}}
    \subfigure[]{
    \includegraphics[width=3.in]{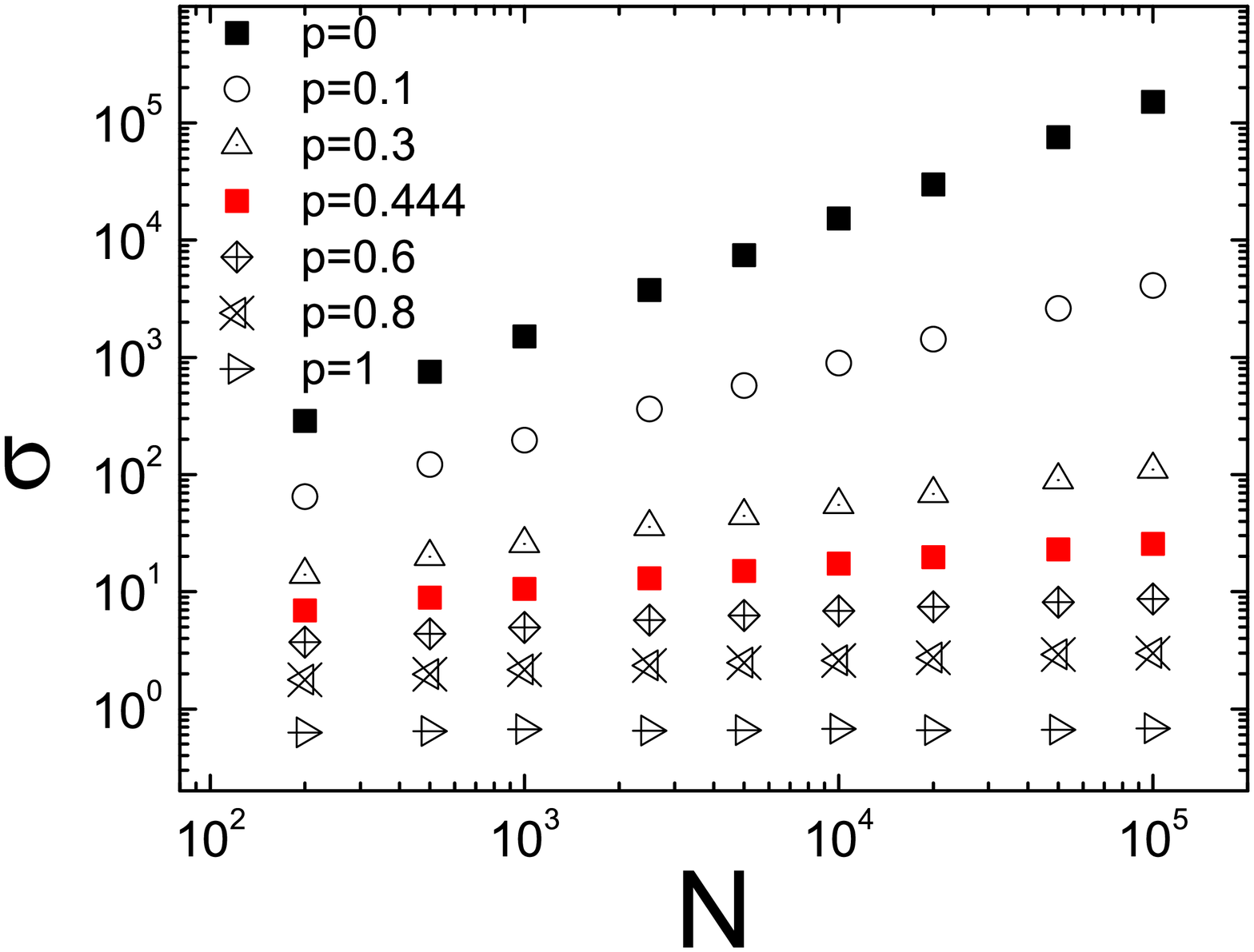}
    \label{fig4:c}}
    \subfigure[]{
    \includegraphics[width=3.in]{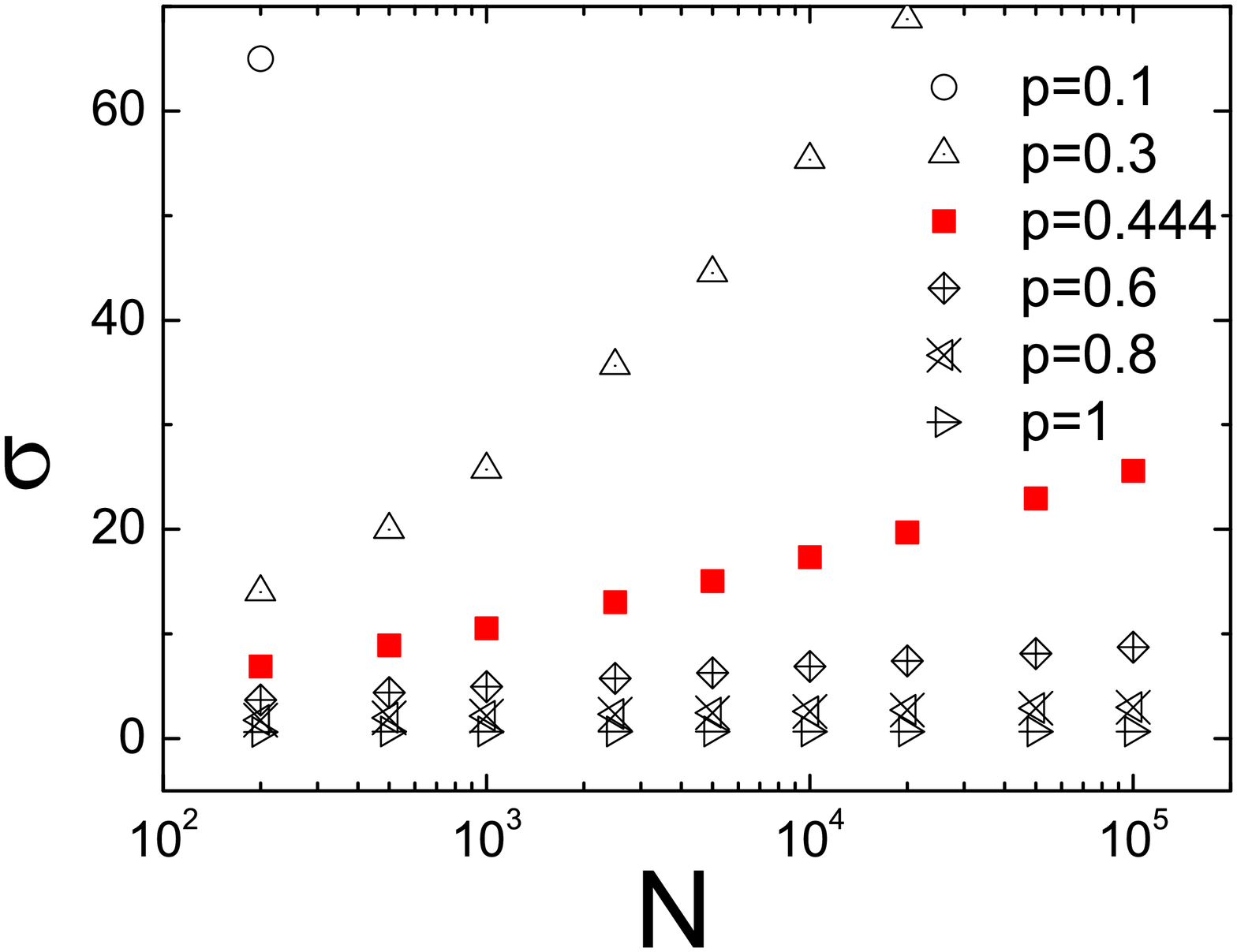}
    \label{fig4:d}}
    \end{center}
     \caption{(a) Log-log and (b) Log-linear plot of MFPT ($T_D$) vs $N$, and (c) Log-log and (d) Log-linear plot of $\sigma$ vs $N$ for ER networks with $\av{k}=10$ for various values of $p$. Symbols are from simulations and lines are from theory.}
\label{fig4}
\end{figure}

We now investigate the case of ER networks. A notable result is the fact that the MFPT in ER networks behaves in the same way as in RR networks i.e. the previous analytical relations are also applicable for ER networks by simply substituing $k$ by $\av{k}$. In fact, in Fig. \ref{fig4:a} and \ref{fig4:b} we see that for ER networks there is also a very good agreement between theoretical and  computed results, and the threshold value of the bias parameter is given now by $p_{th}=(\av{k}-2)/(2\av{k}-2)$. This is an important result since ER networks constitute a more general ensemble than RR networks. In Fig. \ref{fig4:c} and \ref{fig4:d} we see the two regimes of the scaling of the standard deviation $\sigma$  of the FPT for ER networks.

In summary, a model was developed to study the efficiency of biased random walks in networks. The bias is expressed by the parameter $p$, which is the probability that the particle remains in the shortest path to a target node, in the range of extreme values  0 (unbiased case) and 1 (fully biased case).  In both RR and ER networks, the MFPT scaling with the size of the system shows a sudden transition from power law to logarithmic behavior and this transition occurs for the value of the bias parameter $p_{th}=(\av{k}-2)/(2\av{k}-2)$. This was shown by means of Monte Carlo simulations, but also demonstrated analytically with an exact solution for the case of RR networks. Also, a similar transition between two regimes is observed for the standard deviation.

\begin{acknowledgments}
Aknowledgements: This research has been co-financed by the European Union (European Social Fund - ESF) and Greek national funds through the Operational Program "Education and Lifelong Learning" of the National Strategic Reference Framework (NSRF) - Research Funding Program: Heracleitus II (to LS). SH wishes to thank the LINC and the Epiwork EU projects, the DFG and the Israel Science Foundation for support.
\end{acknowledgments}

\clearpage

\bibliography{netbias3}

\end{document}